\documentclass[%
amsmath,amssymb,prl,superscriptaddress,twocolumn
]{revtex4-2}

\usepackage{graphicx}
\usepackage{dcolumn}
\usepackage{bm}
\usepackage{hyperref}
\usepackage{natbib}
\usepackage{float}
\usepackage{amsmath}
\usepackage{xcolor}

\usepackage[utf8]{inputenc}
\DeclareUnicodeCharacter{2212}{\textminus}
\begin{document}

\title{Ubiquitous missing first Shapiro step in Al-InSb nanosheet Josephson junctions}

\author{Xingjun Wu}
\affiliation{Beijing Academy of Quantum Information Sciences, Beijing 100193, China}

\author{Haitian Su}
\affiliation{Beijing Key Laboratory of Quantum Devices and School of Electronics, Peking University, Beijing 100871, China}
\affiliation{Institute of Condensed Matter and Material Physics, School of Physics, Peking University, Beijing 100871, China}

\author{Chuanchang Zeng}
\affiliation{Beijing Academy of Quantum Information Sciences, Beijing 100193, China}

\author{Ji-Yin Wang}
\affiliation{Beijing Academy of Quantum Information Sciences, Beijing 100193, China}

\author{Shili Yan}
\affiliation{Beijing Academy of Quantum Information Sciences, Beijing 100193, China}

\author{Dong Pan}
\email{pandong@semi.ac.cn}
\affiliation{State Key Laboratory of Superlattices and Microstructures, Institute of Semiconductors,Chinese Academy of Sciences, P.O. Box 912, Beijing 100083, China}

\author{Jianhua Zhao}
\affiliation{State Key Laboratory of Superlattices and Microstructures, Institute of Semiconductors,Chinese Academy of Sciences, P.O. Box 912, Beijing 100083, China}

\author{Po Zhang}
\email{zhangpo@baqis.ac.cn}
\affiliation{Beijing Academy of Quantum Information Sciences, Beijing 100193, China}

\author{H. Q. Xu}
\email{hqxu@pku.edu.cn}
\affiliation{Beijing Key Laboratory of Quantum Devices and School of Electronics, Peking University, Beijing 100871, China}
\affiliation{Beijing Academy of Quantum Information Sciences, Beijing 100193, China}

\begin{abstract}
The absence of odd-order Shapiro steps is a predicted signature of topological superconductors. Experimentally, the missing first-order Shapiro step has been reported in both putative topological superconducting systems and topologically trivial superconductor-semiconductor Josephson junctions. Here, we revisit this phenomenon in topologically trivial Al-InSb nanosheet Josephson junctions under microwave irradiation. The missing first Shapiro step coincides with a sharp voltage jump during superconducting switching, yet reappears when the jump is lowered and softened by increasing microwave power, temperature, or magnetic field. It also reappears at higher microwave frequencies, consistent with qualitative results from an RSJ model incorporating the sharp jump. These observations indicate that the absence of the first Shapiro step, associated with the sharp switching jump, simply results from their location within the measurement blind region. This work identifies a common but overlooked mechanism underlying the missing first Shapiro step, offering new insights into fractional Josephson effect experiments.
\end{abstract}

\maketitle

\begin{figure*}
\includegraphics[width=1\linewidth]{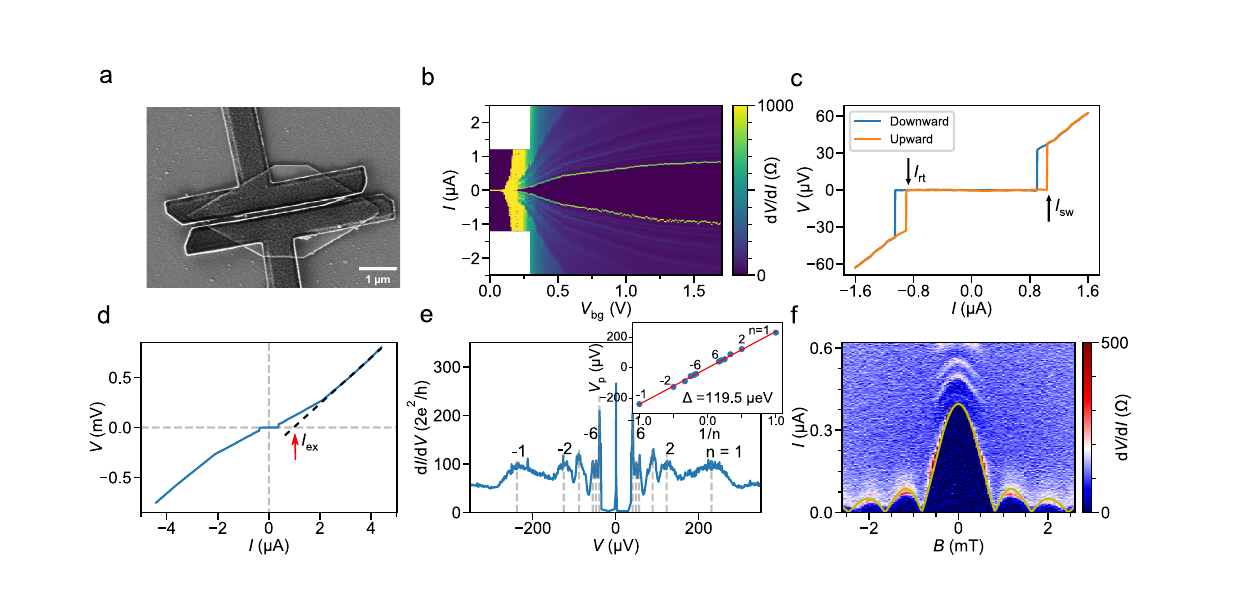}
\caption{Device A characterization. 
(a) Scanning electron microscopy image of device A. The InSb nanosheet (bright gray) is in a hexagonal shape. Two T-shaped superconducting leads (Ti/Al, dark gray) are above the nanosheet. The metallic backgate (Ti/Au) is underneath the nanosheet and extends beyond the displayed range.
(b) Differential resistance d$V$/d$I$ as a function of bias current $I$ and backgate voltage $V_{bg}$. 
(c) Hysteresis in $V-I$ curves scanned in the downward (blue) and upward (orange) directions. Retrapping ($I_{rt}$) and switching ($I_{sw}$) currents are indicated for the upward scanned curve. $V_{bg} = 2$~V. 
(d) Voltage-current characteristic at $V_{bg} = 0.5$~V. The dashed fitting line extrapolates to a finite excess current $I_{ex}$ at $V = 0$ (red arrow). 
(e) Differential conductance d$I$/d$V$ as a function of voltage bias $V$. The vertical dashed lines indicate peak positions due to multiple Andreev reflections, which are used for the fitting (red line) in the inset. An induced gap of 119.5~$\mu$eV is extracted. $V_{bg}$ = 0.5~V. 
(f) Differential resistance d$V$/d$I$ as a function of $I$ and the magnetic field $B$. The yellow line is the switching current fitted with the theoretical Fraunhofer curve. $V_{bg} = 0.5$~V.}
\label{fig_overview}
\end{figure*}

\begin{figure*}
\includegraphics[width=1\linewidth]{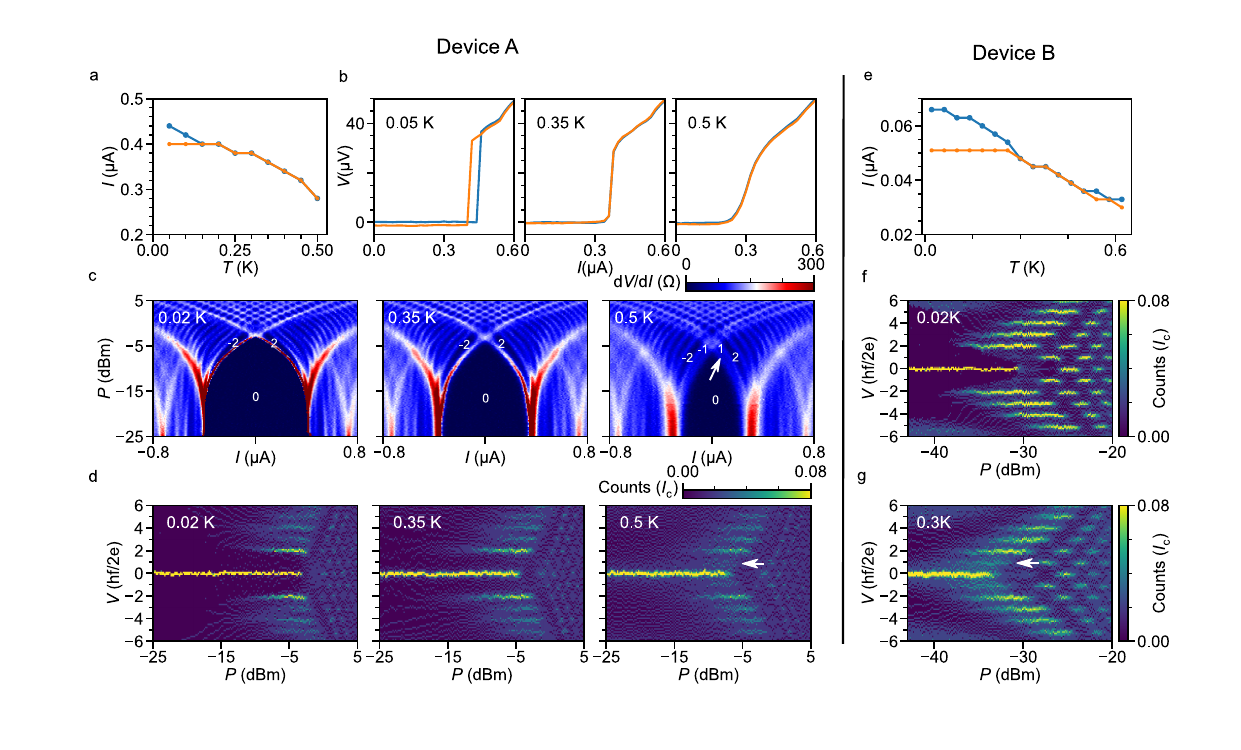}
\caption{Temperature dependence of the missing first Shapiro step and the sharp switching jump. (a)-(d) are from device A. (e)-(g) are from device B.
(a) The switching current (blue) and retrapping current (orange) as a function of the temperature $T$.
(b) Voltage-current characteristics at $0.05$, $0.35$, and $0.5$~K without microwave irradiation. The blue (orange) curve is scanned in the positive (negative) direction. The sharp superconducting switching jump softens as the temperature increases.
(c) Differential resistance $dV/dI$ as a function of microwave power $P$ and DC bias current $I$ at $0.02$, $0.35$, and $0.5$~K. 
(d) Corresponding voltage histograms of (c).
(e) Switching current (blue) and retrapping current (orange) from device B.
(f) and (g) Histogram maps of device B at 0.02 and 0.3~K, respectively.
White arrows in (c), (d), and (g) indicate reappearance of the missing first Shapiro step at higher temperatures. The microwave frequency is 2.5~GHz for both devices. The bin size for all histogram maps is 0.14.
}
\label{fig_temperature}
\end{figure*} 

\begin{figure*}
\includegraphics[width=1\linewidth]{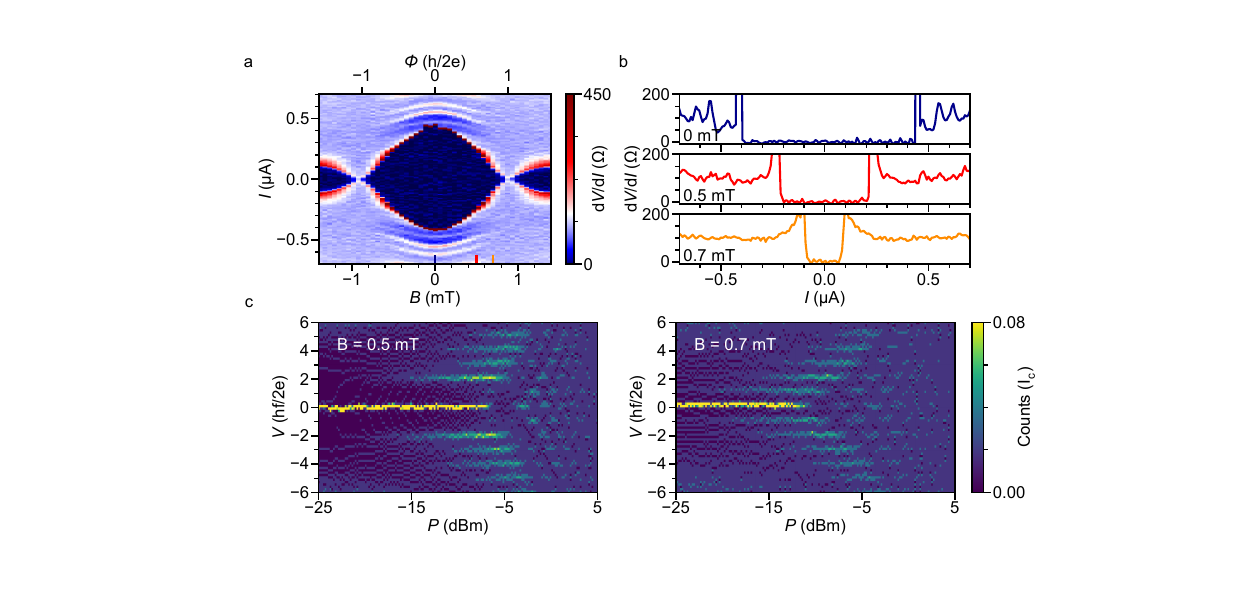}
\caption{Magnetic field dependence of the missing first Shapiro step in device A. 
(a) Differential resistance d$V$/d$I$ as a function of current $I$ and magnetic field $B$, without microwave irradiation. Three short vertical lines near the bottom axis indicate magnetic fields where curves in (b) are extracted.
(b) Linecuts from (a) taken at fixed magnetic fields. The magnetic fields are indicated in each subpanel. The d$V$/d$I$ peaks near superconducting switchings are broadened by the magnetic field. 
(c) Voltage histograms taken at a microwave frequency of 2.5~GHz. The magnetic field is indicated in white text. The bin size is 0.17.
The zero field histogram can be found in Fig.~\ref{fig_temperature}d.}
\label{fig_field}
\end{figure*}

\begin{figure*}
\includegraphics[width=1\linewidth]{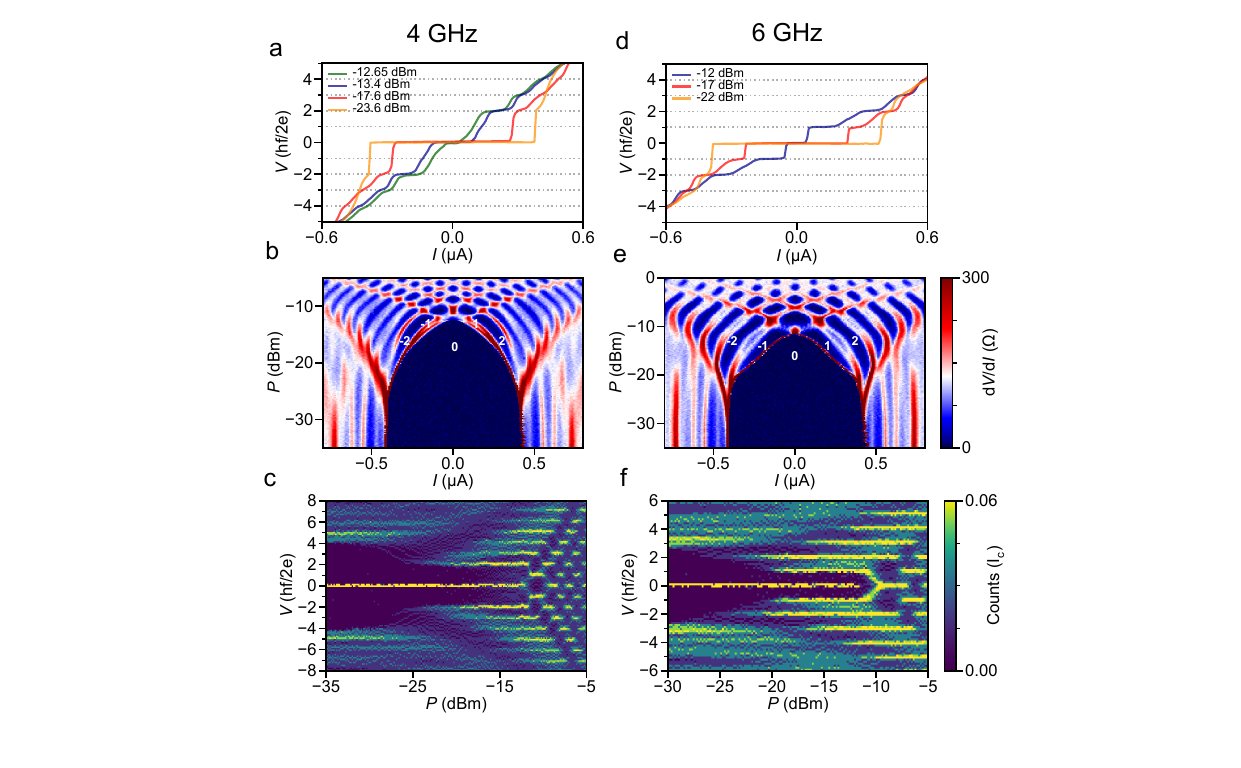}
\caption{Frequency dependence of the missing first Shapiro step in device A.
(a) Voltage-current characteristics at a variety of microwave powers at 4~GHz. The voltage is normalized by $hf/2e$, thus the value corresponds to the Shapiro index.
(b) Differential resistance dV/dI as a function of microwave power $P$ and current $I$  at a microwave frequency of 4~GHz. Shapiro step indexes are indicated in white text.
(c) Voltage histograms that correspond to (b). The bin size is 0.14.
(d)-(f) Similar to (a)-(c) but measured at a frequency of 6~GHz.
The 2.5~GHz response can be found in Fig.~\ref{fig_temperature}.
}
\label{fig_frequency}
\end{figure*}

Topological materials with special energy band structures have received widespread attention in the past few decades~\cite{thouless1982prl, niu1985prb, halperin1993prb, A.Yu.Kitaev_2001, Kane2005prl, bernevig2006quantum, Fu2007prl, lutchyn2010majorana, oreg2010helical}.
Experimental discoveries of new topological states rely on a series of unique signatures, termed ``smoking gun" signatures. Examples of smoking gun signatures for topological materials include quantized conductance in transport experiments~\cite{klitzing1980prl,tsui1982prl, konig2007science, knez2011prl, chang2013experimental} and distinctive energy spectrum patterns in the angle-resolved photoemission spectroscopy~\cite{hsieh2008topological, xu2015science, lv2015prx}. While these signatures have enabled significant progress, their origins can be complex and less straightforward in systems such as topological superconductors~\cite{chen2019prl, kayyalha2020science, yu2021np, dartiailh2021nc, valentini2021science, valentini2022majorana, sato2022prl, wang2022prb, frolov2023smoking}. Controversial conclusions may emerge due to the complexity of real mesoscopic devices compared to theoretical models, potential biases in the search for target phenomena, and an incomplete understanding of new signatures. To determine whether a new topological phase has been achieved and to clarify the physics behind experimental signatures, it is essential to explore these signatures in more deterministic systems.


Topological superconductors hosting Majorana zero modes hold potential for realizing physically protected topological qubits~\cite{read2000paired, A.Yu.Kitaev_2001, kitaev2003fault, nayak2008nonabelian, fu2008prl, oreg2010helical, lutchyn2010majorana}. While such qubits remain undeveloped, the search for Majorana zero modes has spurred extensive research into related experimental signatures~\cite{mourik2012science, deng2012anomalous, das2012zero, rokhinson2012np, nadj2014observation, deng2016majorana, sun2016majorana, Bocquillon2017nn, wang2018evidence, Li2018nm, wang2018prl, rosen2021fractional}. The simplest and the most famous signatures in transport experiments are the zero-bias conductance peak and the missing Shapiro steps. In this work, we focus on the missing Shapiro step signature.

The missing odd-order Shapiro steps, or the fractional AC Josephson effect, can be understood as follows: In microwave-illuminated conventional Josephson junctions (JJs), quantized voltage plateaus (Shapiro steps) emerge from coherent transport of charge $2e$ Cooper pairs, where $e$ is the elementary charge~\cite{josephson1962possible}. These steps occur at $n h f/2e$, where $n = 1, 2, 3, ..$, $h$ is the Planck constant, and $f$ is the microwave frequency. In JJs consisting of two topological superconductors, quasiparticles with half the charge of a Cooper pair is transferred, leading to a doubling of Shapiro step voltages~\cite{Kwon2004epjb, fu2009prb, jiang2011prl, Badiane2011PRL, sanjose2012prl, Dominguez2012prb, Houzet2013PRL, Sau2017prb, li2018prb}. The fractional AC Josephson effect can also be understood with the periodicity of a JJ's current-phase relation (CPR). A topological JJ has a $4\pi$-periodic CPR, in contrast to the $2\pi$-periodic CPR for conventional JJs. The doubling in periodicity leads to the doubling of Shapiro step voltages, i.e., the absence of odd-order steps~\cite{Park2021prb}. Overlapping of Majorana zero modes in a finite system introduces a gap in the energy spectrum and brings a $2\pi$ Josephson component. However, the $4\pi$ term persists due to topological protection~\cite{pikulin2012prb}. Beyond topological mechanisms, topologically trivial processes\textemdash including Landau-Zener transitions~\cite{sau2012possibility, Sau2017prb}, non-constant resistance~\cite{mudi2022model}, and Joule heating (which can suppress a continuous series of Shapiro steps starting from the first)~\cite{Cecco2016prb, Shelly2020pra}\textemdash have also been proposed to explain missing odd-order Shapiro steps.


Experimentally, the missing first Shapiro step is reported in a variety of materials hypothesized to be topologically non-trivial, including hybrid junctions combining s-wave superconductors and semiconductor nanowires, topological insulators, or
Dirac semimetals~\cite{rokhinson2012np, wiedenmann2016nc, Li2018nm, yu2018prl, bai2022proximity, wang2018prl, rosen2021fractional}. A stronger indication\textemdash with a series of missing odd-order Shapiro steps\textemdash has been reported exclusively in the Al-HgTe 2D topological insulator system~\cite{Bocquillon2017nn}.  In topologically trivial regimes, experiments on Al-InAs quantum well hybrid systems under zero magnetic field have reported both missing first~\cite{dartiailh2021nc} and multiple odd-order Shapiro steps~\cite{zhang2022missing}. 

While both theoretical and experimental studies have indicated that missing odd-order Shapiro steps\textemdash particularly the missing first Shapiro step\textemdash can arise from non-topological origins, the underlying mechanisms and their prevalence in experimental settings, as we explore in this work, remain as important open questions, requiring further investigation.

\section{List of key results}

We study Shapiro steps in Al-InSb nanosheet-based JJs. Instead of growing both materials in the same molecular beam epitaxy (MBE) machine, the Al layer is grown ex-situ using conventional nanofabrication methods. The magnetic field is either set to be zero or the order of millitesla, much smaller than the field required for a topological transition~\cite{Dartiailh2021prl}. Therefore, we study the system in the topologically trivial regime. In the absence of microwave irradiation, the voltage-current characteristics exhibit a sharp voltage jump near the superconducting switching transition. Elevating temperature, microwave power or applying a magnetic field lowers and softens this jump, prompting the reappearance of the previously missing first Shapiro step. Notably, our experiments reveal no direct correlation between the presence/absence of the first Shapiro step and the junction's hysteretic behavior. Sharp voltage jumps can persist in non-hysteretic devices, similarly inducing the disappearance of the first Shapiro step. In addition, the missing first Shapiro step reappears with an increased microwave frequency, which can be qualitatively captured by introducing a sharp voltage jump into the RSJ model. Beyond the missing first Shapiro step, we also observe a residual supercurrent at the first node of the zeroth Shapiro lobes and a suppression of the third Shapiro step.

\section{Methods}

High-quality free-standing InSb nanosheets are grown by MBE. 
These nanosheets typically have dimensions on the order of micrometers in length and width, 10 to 100~nm in thickness. The low-temperature mobility is above $10^4$~cm$^2$/Vs. Details about the material and transport properties can be found in Refs.~\cite{pan2016nl, Kang2019nl,Wu_2025}. Two devices are studied in the main text, termed devices A and B. Nanosheets are transferred to the substrate by a micromanipulator. Local Ti/Au backgates are pre-patterned before depositing nanosheets. Native oxide on the surface of the nanosheet is removed by sulfur passivation~\cite{Suyatin_2007, Kang2019nl}. The superconducting leads are patterned by e-beam lithography followed by evaporating 5/110 nm Ti/Al. The junction width and length are 4.2 (1.9)~$\mu$m and 110 (120)~nm in device A (B). 

Measurements are performed in a dilution refrigerator with a base temperature of about 30~mK. The microwave is coupled to junctions via an antenna. Attenuators of 29~dB in total are installed in the microwave line~\cite{Yan2023}. The power value marked in the manuscript is the power on the microwave source. A numeric offset of about 3.5~mT is applied to the magnetic field. The offset may arise due to fluxes trapped in the magnet. The voltage is shifted by values of the order of $\mu$V to compensate for DC offsets in amplifiers and thermal voltages in lines.

\section{Figure~\ref{fig_overview} description}

The characterization of device A is presented in Fig.~\ref{fig_overview}. Two T-shaped Ti/Al leads and the InSb in between form a JJ (Fig.~\ref{fig_overview}a). Superconductivity of the junction can be tuned by the metallic backgate underneath the nanosheet. The turn-on backgate voltage is about 0.2~V (Fig.~\ref{fig_overview}b). The switching/retrapping currents reach $-1$/$0.8$~$\mu$A as $V_{bg}$ increases to 1.7~V. Dark fringes alongside the central zero-resistance lobe are due to multiple Andreev reflections (MARs). The difference between switching and retrapping currents indicates a hysteresis regarding the current, which is shown more clearly in Fig.~\ref{fig_overview}c.

The typical excess current $I_{ex}$ and normal-state resistance $R_n$ are extracted from the fit to the linear region in Fig.~\ref{fig_overview}d, which gives $I_{ex} = 0.97$~$\mu$A, $R_n = 230$~$\Omega$, and an $I_{ex} R_n$ product of 223~$\mu$V. We estimate an induced superconducting gap $\Delta$ of 119.5~$\mu$eV by fitting positions of MARs to $V = 2 \Delta/n e$, $n = \pm 1, \pm 2, ..., \pm 6$ (Fig.~\ref{fig_overview}e). The induced gap is roughly half of the value reported in devices with epitaxial Al layer~\cite{zhang2022prl}. The small induced gap is expected because, in this work, the Al layer is grown using conventional nanofabrication methods. Despite a smaller gap, the device also manifests high-quality coherent transport properties, demonstrated by MARs up to an order of six. This is likely because the Andreev reflection does not take place at the physical superconductor-semiconductor interface
and the high mobility in the junction's normal region~\cite{Shabani2016prb}. We extract a high transparency $T_{r} $ of 0.9 using the Octavio-Tinkham-Blonder-Klapwijk (OTBK) model~\cite{PhysRevB.27.6739}, given that $eI_{ex}R_{n}/\Delta = 1.86$.

In magnetic fields, the switching current oscillates following the Fraunhofer diffraction pattern (Fig.~\ref{fig_overview}f), indicating a uniform current distribution in the device. The period 0.8~mT gives an effective junction length of about 600~nm, which is larger than the designed junction length (110~nm). The larger effective length is likely due to the London penetration or flux focusing~\cite{Kang2019nl, zhang2024sp}.

\section{Figure~\ref{fig_temperature} description}

Temperature dependence of the missing first Shapiro step is presented in Fig.~\ref{fig_temperature}. In the absence of microwave irradiation, the extracted switching and retrapping current curves merge at 0.15~K (Fig.~\ref{fig_temperature}a), consistent with the explanation of Joule heating of the hysteresis~\cite{Courtois2008prl}.  In addition to the hysteresis, sharp superconducting switching jumps are observed in the voltage-current characteristics (Fig.~\ref{fig_temperature}b). At 0.05~K, the curve exhibits both sharp switchings and hysteresis. For example, in the upward scan (blue), the voltage undergoes an abrupt jump from $0$~V at 0.44~$\mu$A to 37~$\mu$V at 0.46~$\mu$A.  This sharp jump persists at 0.35~K, whereas the hysteresis is no longer discernible at this temperature, suggesting that the jump is independent of the hysteresis. At 0.5~K, the switching becomes significantly softer, with a smaller slope in the vicinity of the switching current. The softening of the switching is further corroborated by the broadening of $dV/dI$ peaks at -25~dBm in Fig.~\ref{fig_temperature}c.

Next, we turn on the microwave irradiation and study Shapiro steps. A Shapiro step is a minimum in the $dV/dI$ maps (Fig.~\ref{fig_temperature}c) or a maximum in the histograms (Fig.~\ref{fig_temperature}d). We are interested in Shapiro steps at power values smaller than the first node of the zeroth Shapiro step lobe~\cite{Bocquillon2017nn, dartiailh2021nc}. The first Shapiro step is missing at 0.02 and 0.35~K. Furthermore, at 0.02~K the sharp switching leaves a dark region between $V = 0$ and $2$ (Fig.~\ref{fig_temperature}d), meaning that there are no data points collected in this voltage range due to the sharp switching. In contrast, the region between $V=2$ and 3 is relatively brighter. At 0.35~K, the area between $V = 0$ and $2$ becomes brighter due to a slight softening of the switching. At 0.5~K, although thermal fluctuations obscure the Shapiro map, the reappearance of the first Shapiro step is still visible. This can be confirmed by looking at either the splitting of $dV/dI$ peaks in Fig.~\ref{fig_temperature}c (white arrow) or the voltage histogram in Fig.~\ref{fig_temperature}d (white arrow).

Similar behaviors are observed in device B (Figs.~\ref{fig_temperature}e-\ref{fig_temperature}g). Unlike in device A, the missing first Shapiro step in device B reappears as soon as the hysteresis disappears at 0.3~K. We do not observe an intermediate regime where the hysteresis disappears while the first Shapiro step is still missing.

\section{Figure~\ref{fig_field} description}

The magnetic-field dependence of the missing Shapiro step in device A is depicted in Fig.~\ref{fig_field}. We first study the response without microwave irradiation. Fig.~\ref{fig_field}a shows a zoomed-in view of Fig.~\ref{fig_overview}f near zero field. Vertical cuts along $B = 0$, 0.5, and 0.7~mT are extracted and depicted in Fig.~\ref{fig_field}b. Similar to the temperature, the magnetic field not only reduces the critical current but also widens the $dV/dI$ peaks at the superconducting switching. The broadening of the $dV/dI$ peaks above 0.5~mT can be observed in both Figs.~\ref{fig_field}a and \ref{fig_field}b. 

Under microwave irradiation, the first Shapiro step reappears at 0.7~mT (Fig.~\ref{fig_field}c). In the intermediate regime where the field is $0.5$~mT, the onset of the first missing Shapiro step occurs near -5.6~dBm. In the histogram, there are still dark regions with small bin counts near $V = \pm 1$ and $P < -7$~dBm, indicating that the switching is still sharp. At 0.7~mT, the same area is brighter, and the stripe of the first Shapiro step is observed.

Another interesting effect is the suppression of higher odd-order Shapiro steps, here the third Shapiro step. The onset power of the third step is larger than that of the two adjacent steps at 0 and 0.5~mT. As a comparison, the onset power increases monotonically with the Shapiro index at 0.7~mT. Similar suppression is reported in the trivial regime of the Al-InAs quantum well system~\cite{dartiailh2021nc, zhang2022missing}. The non-topological origin of the suppression could be either Landau-Zener transition or non-linear resistance due to MARs~\cite{mudi2022model}. The latter explanation is favored in this work because the suppression disappears once the MARs are smoothed out by increasing the magnetic field to 0.7~mT (Fig.~\ref{fig_field}a and \ref{fig_field}c).

\section{Figure~\ref{fig_frequency} description}

We study the frequency dependence of the missing Shapiro steps in Fig.~\ref{fig_frequency}. At an increased microwave frequency of 4~GHz (Figs.~\ref{fig_frequency}a-\ref{fig_frequency}c), the $dV/dI$ peak between dips corresponding to steps $0$ and $2$ splits. The splitting indicates the onset of the first Shapiro step (labeled ``1") that is not visible at $2.5$~GHz. The first Shapiro step is still indiscernible in the corresponding histogram (Fig.~\ref{fig_frequency}c) because the step is not well quantized. The bin counts within the region where $P < -15$~dBm and $0 < V < 2$ are close to zero, resulting in a dark appearance for this area. Below -25 dBm, the dark area extends to $V = 4$, ``eliminating" Shapiro steps 1, 2, and 3. This is another piece of evidence that indicates missing Shapiro steps are caused by the sharp switching jump. 


At a higher frequency of 6~GHz (Figs.~\ref{fig_frequency}d-\ref{fig_frequency}f), the first step is clearly established. In addition to the reappearance of the first step,  we observe half-integer steps at 6~GHz. The most apparent half step is the 3/2 step, which is a dip in $dV/dI$ between dark blue areas labeled ``1" and ``2" or the bright stripe at $V = 1.5$ in the histogram. Higher-order half-integer steps are also observed. However, the first half-integer step, the ``1/2" step, is missing. Similar results are reported in the quantum well system~\cite{zhang2022missing}. 
We attribute it also to the sharp switching in the V-I curve, evidenced by the near zero bin counts at $V = \pm 1/2$.

A residual supercurrent at the node of the zeroth Shapiro step lobe is observed in Fig.~\ref{fig_frequency}b. A zoomed-in view of this figure and residual supercurrent in device B can be found in the supplemental material~\cite{suppmat}. The residual supercurrent is also a signature for topological superconductors~\cite{le2019joule, wang2018prl}. However, higher-order Josephson harmonics in the CPR of a JJ can also mimic this feature~\cite{suppmat}.

\section{Discussion}

Previous works on the missing Shapiro steps were carried out in systems hypothesized to be topologically non-trivial or in the non-topological regime of the Al-InAs quantum well system~\cite{dartiailh2021nc, zhang2022missing}. The latter has superconducting layers grown using dedicated methods combining MBE, in-situ growth or hydrogen plasma cleaning, and cryogenic temperature growth techniques, which are not used in many topological experiments~\cite{Shabani2016prb, Lee2019nl, zhang2022planar}. These techniques guarantee a sharp superconductor-semiconductor interface and the uniformity of the superconducting layer~\cite{Krogstrup2015nm,Pan2022cpl}. Devices fabricated via conventional nanofabrication methods in this work exhibit an induced gap approximately half the typical size of that in epitaxial Al devices; this is attributed to the imperfect interface between the superconductor and semiconductor regions~\cite{PhysRevApplied.7.034029,PhysRevB.55.8457,PhysRevB.53.365,VOLKOV1995261}. Since many works regarding the missing Shapiro steps do not employ the epitaxial method, devices in this work serve as a better comparison platform for topological superconductor studies. The observation of the missing first Shapiro and other signatures in this work reveals that these signatures are unexpectedly common in experiments.

Sharp switching behavior is a common phenomenon in JJ systems\cite{PhysRevLett.77.3435,Courtois2008prl,PhysRevApplied.22.064027,PhysRevLett.85.170,bretheau2013exciting,PhysRevApplied.22.064035}, which can be equivalently interpreted through the dynamics of a phase particle in a tilted washboard potential. For a small bias current $I < I_{c}$, the particle is trapped in a potential well, corresponding to a metastable zero-voltage state. When the bias current exceeds the critical current ($I > I_{c}$), the particle escapes the well and accelerates down the washboard, switching to a resistive state. In the range of $0 < I < I_{c}$, the particle can hop diffusively between wells via thermal and quantum fluctuations, inducing stochastic switching events. Such stochastic events are closely related to the junction dynamics. By analyzing the switching probability distributions as a function of bias current, we infer that the device operates in the macroscopic quantum tunneling regime at low temperatures~\cite{suppmat}. The sharp switching behavior gives rise to a measurement "blind" region, which underlies the disappearance of Shapiro steps (see Supplemental Material~\cite{suppmat} for details). 

Several pieces of evidence indicate that the missing Shapiro steps\textemdash including the first step\textemdash stem from such sharp superconducting switching. 
First, the bin counts are close to zero at $V = \pm 1$ when the first step is missing, indicating a violent voltage change; notably, the first step reappears as the switching jump softens. 
Second, the disappearance of the $1/2$ Shapiro step, and the elimination of all Shapiro steps below $V = 4$ by the extension of the dark region in histograms. These missing steps can be understood in the same way as the disappearance of the first Shapiro step. In contrast, Majorana physics or Landau-Zener transition fails to include these missing steps. Third, the frequency dependence favors the sharp switching jump scenario over the Landau-Zener transition scenario. The former mechanism is enhanced with a decreased frequency because the lower the frequency, the smaller and more fragile the Shapiro step voltages compared to the fixed jump voltage. The latter mechanism, however, is weakened with a decreased frequency. The missing first Shapiro step, the frequency dependence, and the extension of the dark region at lower powers are qualitatively captured by a conceptional model where the voltage jump is introduced by adding a shunted capacitance~\cite{suppmat}.

In conclusion, the observation of the missing first Shapiro step, together with other signatures, in the topologically trivial regime of Al-InSb nanosheet JJs reveals the unexpected ubiquity of these signatures. The missing first Shapiro step is attributed to sharp superconducting switchings in voltage-current characteristics taken without microwave irradiation, a common but not well-recognized phenomenon in real devices. Even though the experiment is conducted in a preset topologically trivial regime, the analysis employed in this study does not depend on any prior knowledge regarding the system's topological properties. Therefore, it can be employed in demonstrating the presence of topological states in other materials. 
Future Majorana works may focus on improving the quality and tunability of devices. The smoking gun signature paradigm should still work by deeper analysis of larger volumes of data, including data from comparison experiments made with topologically trivial systems~\cite{frolov2023smoking}.

\section{Data availability}

Data, processing code, and simulation code are available at Ref.~\cite{zenodo}

\section{Acknowledgements}

This work is supported by the NSFC (Grant Nos. 92165208, 12004039, 11874071, 92365103, 12374480, 12374459, 61974138 and 92065106). D.P. acknowledges the support from Youth Innovation Promotion Association, Chinese Academy of Sciences (Nos. 2017156 and Y2021043). 

\section{Author contributions}

H.Q.X. supervised the project. X.W., H.S., and S.Y. fabricated devices. X.W., H.S., and J.-Y.W. performed measurements. C.Z. and P.Z. did the simulation. D.P. and J.Z developed the nanosheet material. X.W., P.Z., and H.Q.X. analyzed the data and wrote the manuscript with inputs from all authors.

\bibliography{ref.bib}

\end{document}